\documentclass[aps,twocolumn,showpacs,superscriptaddress,groupedaddress]{revtex4}  
\usepackage{amsmath,amssymb}
\usepackage{hyperref}
\usepackage{graphics,psfrag}
\usepackage{graphicx,psfrag}
\usepackage{longtable}
\usepackage{tabularx}
\usepackage{booktabs}
\usepackage[utf8x]{inputenc}
\usepackage[USenglish]{babel}
\usepackage{multirow}
\usepackage{rotating}
\usepackage{hhline}
\usepackage{xcolor}
\usepackage{ulem}
\usepackage{todonotes}
\usepackage{array}
\usepackage{nicefrac}
\usepackage[hang]{subfigure}
\usepackage{soul}

\begin{document}

\title{DFT+U simulation of the Ti${}_4$O${}_7$-TiO${}_2$ interface}

\author{A. C. M. Padilha}
\email{antonio.padilha@ufabc.edu.br}
\affiliation{Centro de Ciências Naturais e Humanas, Universidade Federal do ABC, Santo André, SP, Brazil}

\author{A. R. Rocha}
\affiliation{Instituto de Física Teórica, Universidade Estadual Paulista (UNESP), São Paulo, SP, Brazil}

\author{G. M. Dalpian}
\email{gustavo.dalpian@ufabc.edu.br}
\affiliation{Centro de Ciências Naturais e Humanas, Universidade Federal do ABC, Santo André, SP, Brazil}

\date{\today}

\begin{abstract}
  The formation of conducting channels of Ti${}_4$O${}_7$ inside TiO${}_2$-based memristors is believed to be the origin for the change in electric resistivity of these devices. While the properties of the bulk materials are reasonably known, the interface between them has not been studied up to now mostly due to their different crystalline structures. In this work we present a way to match the interfaces between TiO${}_2$ and Ti${}_4$O${}_7$ and subsequently the band offset between these materials is obtained from density functional theory based calculations. The results show that while the valence band is located at the Ti${}_4$O${}_7$, the conduction band is found at the TiO${}_2$ structure, resulting into a type II interface. In this case, the Ti${}_4$O${}_7$ would act as a donor to the TiO${}_2$ matrix.
\end{abstract}


\maketitle

\section{Introduction}

The memristor is an electronic device predicted by L. Chua in 1971\cite{chua_1971,chua_kang_1976} and experimentally obtained a few years ago.\cite{S_Williams_nature_2008} These devices are characterized by distinct resistance states which could be used to store information as they show very fast switching and high retention times.\cite{Jeong2012} 

There are reports of a wide range of materials (particularly metal oxides) which show this memristive effect.\cite{choi2005,Syu2013,Wang2013,Kim2013a,Huang2013,Kim2013b,He2013,Ignatiev2006,Guo2013,Pilch2014} Although the switching mechanism, and the origin of the different resistance states has not been completely elucidated, it is generally believed that the presence of oxygen deficient phases plays a major role in the memristive effect.\cite{Szot2011} Within the possible materials, TiO${}_2$ is one of the most studied ones.\cite{Szot2011,Kim2011,Janotti2010,choi2005,Jeong2008,Kwon2010} Inside memristive devices, Ti${}_4$O${}_7$ channels immersed in the TiO${}_2$-based matrix have been observed, and the formation of such structures is believed to be largely responsible for the change in resistivity.\cite{Kwon2010}

While TiO${}_2$ is known to be a wide gap semiconductor (experimental band gap $E_g \approx 3.1$ eV\cite{Amtout1995}) Ti${}_4$O${}_7$ presents three phases with different electronic transport properties. The low temperature ($T \leq$ 140 K) and intermediate temperature phases (140 K $\leq T \leq$ 150 K) are semiconductors while the high temperature phase ($T \geq$ 150 K) is metallic.\cite{Bartholomew1969,Marezio1971,Marezio1973,LePage1984} The differences between those phases is mainly due to small atomic displacements which do not lead to significant changes in the unit cell.

Despite the large amount of theoretical work on either TiO${}_2$ rutile \cite{Cho2006,Iddir2007,Li2009a,Janotti2010} and Ti${}_4$O${}_7$,\cite{Eyert2004,Leonov2006,Liborio2009,Weissmann2011,Padilha2014a} the band alignment of these two materials is not known up to now, mainly because of the difficulty in building a supercell containing the Ti${}_4$O${}_7$-TiO${}_2$ interface - the two materials present very different unit cells, making the match between them very difficult. This information could be very useful to the understanding of the atomic level mechanism of the memristor, given that the Ti${}_4$O${}_7$-TiO${}_2$ interface spans a great part of the device length after the formation of the conducting channels.

As the total energy is given with respect to an arbitrary reference energy in periodic calculations, a common reference should be determined for all the structures involved in this work.\cite{Ihm1979} In fact, the knowledge of this common reference is necessary to find the offset between the valence band of different materials. The knowledge of this band offset is important, for instance, in determining whether there is charge transfer at the interface between the two systems, the materials doping limits\cite{Zhang2000} as well as other properties. For instance, Scanlon \textit{et al.}\cite{Scanlon2013} determined the band offset between rutile and anatase TiO${}_2$ structures via a cluster approach, providing an understanding about the separation of charge carriers in photocatalysis using the mixed-phase structure.

Our objective in this paper is to study the interface of TiO${}_2$ and one of its oxygen deficient counterparts, Ti${}_4$O${}_7$. For this purpose, we built a supercell containing the Ti${}_4$O${}_7$-TiO${}_2$ interface which was then used for our calculations. In particular we analyze the band alignment, which has been found to be of type II (the conduction band minimum (CBM) is located in one material and valence band maximum (VBM) in the other), and its consequences to the overall electronic structure of the combined system. 
\begin{center}
 \begin{figure*}[t!]
  \begin{subfigure}[ Ti${}_4$O${}_7$]{
   \includegraphics[width=0.47\textwidth]{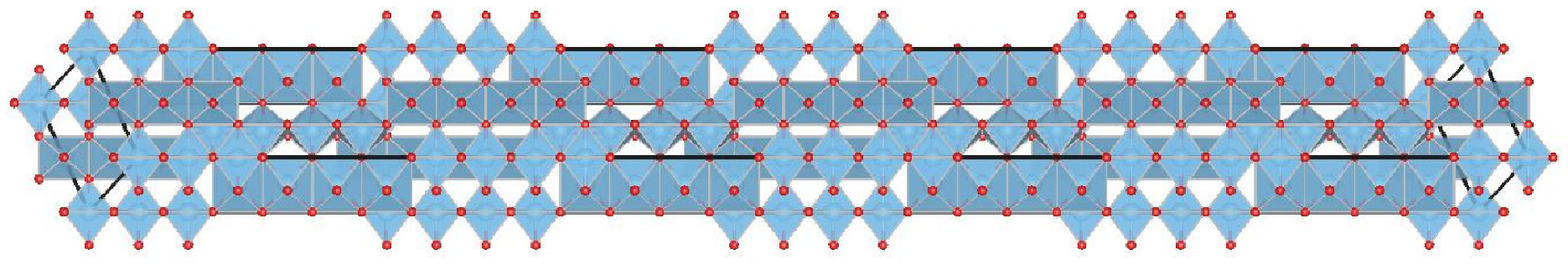}
   \label{fig:struct-ti4o7}
   }
  \end{subfigure}
  \begin{subfigure}[ TiO${}_2$]{
   \includegraphics[width=0.47\textwidth]{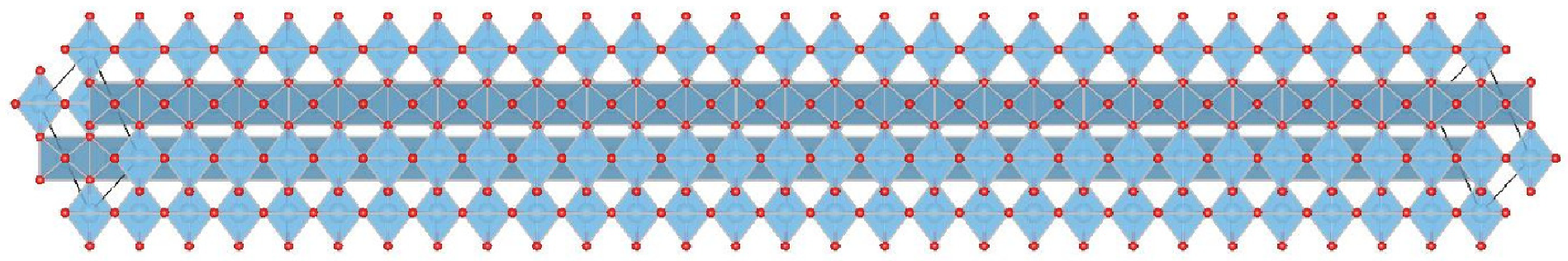}
   \label{fig:struct-tio2}
   }
  \end{subfigure}
  \begin{subfigure}[ Ti${}_4$O${}_7$-TiO${}_2$ interface]{
   \includegraphics[width=0.94\textwidth]{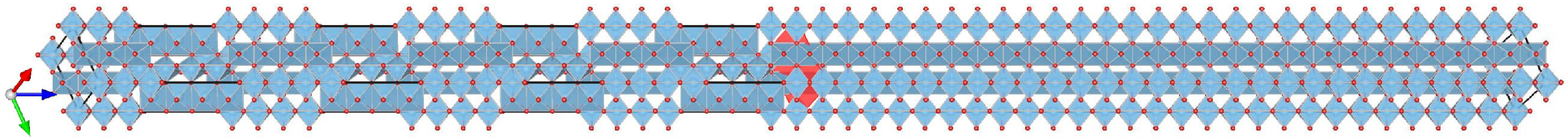}
   \label{fig:struct-interface}
   }
  \end{subfigure}
  \caption{(a) Ti${}_4$O${}_7$ structure generated via the operations given by equations \ref{eq:transf} and \ref{eq:displ}, (b) TiO${}_2$ transformed according to equation \ref{eq:transf}, (c) and Ti${}_4$O${}_7$-TiO${}_2$ interface. The red plane in (c) indicates the interface region and the blue horizontal arrow corresponds to the $c$ crystal vector.}
  \label{fig:struct}
 \end{figure*}
\end{center}

\section{Computational Details}

All density functional theory (DFT\cite{Hohenberg_Kohn_64,Kohn_Sham_65}) calculations in this work were performed using the VASP package.\cite{Kresse1996,KresseFurthmuller1996} Wave functions are described within the Projector Augmented Waves (PAW) scheme\cite{Blochl1994,Kresse1999} with periodic boundary conditions. The $3p3d4s$ and $2s2p$ electrons were considered as valence electrons for Ti and O atoms respectively and the PBE functional\cite{Perdew1997} was used as the exchange-correlation potential. The cutoff energy for the plane wave expansion was 700 eV and the k-point meshes used throughout this work were generated using a $2 \times 2 \times 1$ $\Gamma$-centered grid. Spin polarization was taken into account for all calculations. Relaxation of the ions was performed while the unit cell was kept fixed, until forces were smaller than $2.5 \times 10^{-3}$ eV/\AA, for the isolated materials and $5.0 \times 10^{-2}$ eV/\AA, for the supercell. We have performed test calculations with a smaller supercell for the interface, but with more strict parameters for the convergence of the forces. The results were similar for quantities calculated in this work.

A Hubbard $U$ parameter was introduced for a better description of the Ti(d) electrons. Our earlier work \cite{Padilha2014a} shows that a value of 5 eV for $U$ is enough to achieve a good description of the electronic structure of Ti${}_n$O${}_{2n-1}$ ($2 \leq n \leq 5$) Magnéli phases. Even though, test calculations were performed for other values of $U$ (0, 2 and 4 eV) and the same qualitative results were obtained for the type of alignment as well as the relative position of the unnocupied levels of the same materials. In the rotationally invariant scheme as implemented in VASP\cite{Dudarev} only the value of $U-J$ ($J$ is the spherically averaged matrix element of the screened Coulomb interaction between electrons) is of relevance, thus $J$ was set to zero for all calculations.

\section{Construction of the Interface}

One of the key points in this work is that building the interface requires matching two of the three unit cell vectors of each compound at the interfacial region. For that to be accomplished, it is necessary to find the smallest supercell where the deformation imposed to each compound is such that no significant change in electronic structure with respect to the bulk material arises in regions away from the interface. For the materials presented in this work, this is difficult, since the TiO${}_2$ and the Ti${}_4$O${}_7$ presented very distinct unit cells. The stable form of TiO${}_2$ at ambient conditions is the rutile structure (tetragonal unit cell, $a_r=b_r=4.59$\AA, $c_r=2.96$\AA\cite{Grant1959}) while the Ti${}_4$O${}_7$ presents a complicated Magnéli structure (triclinic cell, see Table \ref{tab:struct}).
\begin{center}
 \begin{table}[h!]
  \centering
   \caption{\label{tab:struct} Cell parameters for Ti${}_4$O${}_7$, high (HT), intermediate (IT) and low temperature (LT) phases\cite{LePage1984}, as well as the result of the transformation given by equation \ref{eq:transf} (TR).}
   \begin{tabularx}{\columnwidth}{*{1}{m{0.1\columnwidth}} *{4}{>{\centering\arraybackslash}m{0.2\columnwidth}}} 
   \hline
   \hline
                                 &  HT         &  IT         &  LT         & TR                 \\ 
    \hline
    $a$ (\AA)                    & 5.60        & 5.60        & 5.63        & 5.46                \\
    $b$ (\AA)                    & 7.12        & 7.13        & 7.20        & 7.14                \\
    $c$ (\AA)                    & 20.43       & 20.36       & 20.26       & 20.70               \\
    $\alpha$ (${}^{\mathrm{o}}$) & 67.70       & 67.76       & 67.90       & 65.52               \\
    $\beta$ (${}^{\mathrm{o}}$)  & 57.16       & 57.44       & 57.69       & 57.22               \\
    $\gamma$ (${}^{\mathrm{o}}$) & 108.76      & 109.02      & 109.68      & 108.46              \\
   \hline
   \hline
   \end{tabularx}
 \end{table}
\end{center}

To overcome this difficulty, we followed steps that lead to the derivation of the Magnéli structure from the rutile one as follows. The first step is to obtain the Magnéli phase structures from TiO${}_2$ rutile via the unit cell transformation\cite{Wood1981}
\begin{equation}
 \begin{array}{cccc}
 \left[ \begin{array}{c} a_M^{(n)} \\ b_M^{(n)} \\ c_M^{(n)} \end{array} \right] & = & \left[ \begin{array}{ccc} -1 & 0 & 1 \\ 1 & -1 & 1 \\ 0 & 0 & 2n-1 \end{array} \right] & \left[ \begin{array}{c} a_r \\ b_r \\ c_r \end{array} \right],
 \end{array}
 \label{eq:transf}
\end{equation}
where $a_M^{(n)}, b_M^{(n)}$ and $c_M^{(n)}$ are the unit cell vectors of the Magnéli structure of index $n$. The result of using this operation for the case when $n=4$ is illustrated in Fig. \ref{fig:struct-tio2} and the comparison with the Ti${}_4$O${}_7$ unit cell is presented in Table \ref{tab:struct}. Notice that this still represents a TiO${}_2$ rutile structure, but the unit cell is now the one from Ti${}_4$O${}_7$. From this point, the Ti${}_4$O${}_7$ Magnéli phase, as well as all Magnéli phases (Ti${}_n$O${}_{2n-1}$, $n = 2,3,\dots,10$) can be generated via the operation\cite{Andersson1960,Harada2010,Anderson19671393}
\begin{equation}
  (121) \frac{1}{2}[0\bar{1}1],
  \label{eq:displ}
\end{equation}
that consists of periodic displacements; after every $n$-th (121) plane, the atoms are translated by $\nicefrac{1}{2}[0\bar{1}1]$ crystal vector of the rutile unit cell. This operation leads to the formation of extended defects, which are composed of oxygen vacancies ($V_O$'s) restricted to planes known as shear planes. The resulting arrangement is presented in Fig. \ref{fig:struct-ti4o7}. Since the operation given by equation \ref{eq:displ} affects only the atomic positions, both structures present the same triclinic unit cell.

In essence, the transformation presented above results in unit cells with equal lengths. The two materials can thus be interfaced in the plane that is secant to the $c$ axis (highlighted in Fig. \ref{fig:struct-interface}). This Ti${}_4$O${}_7$-TiO${}_2$ interface is, by construction, strain free, defect free, presents exactly the same cell parameters and consequently no dipoles would arise at the heterojunction. In our calculations we created a supercell along the $c$ axis by repeating the cell for each material four times ($c=165.61$\AA) in order to ensure convergence of the potential away from the interface.
\begin{center}
 \begin{figure}[Ht!]
  \begin{center}
   \includegraphics[width=\columnwidth]{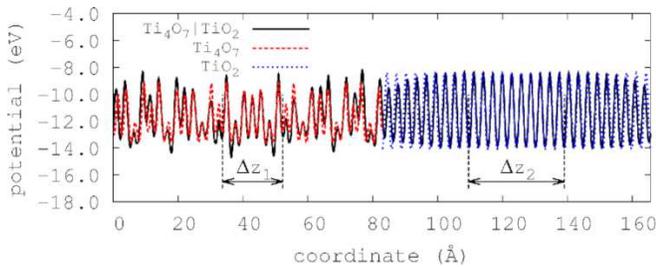}
   \caption{Electrostatic potential profiles along the $c$ vector for all structures presented in this work: isolated TiO${}_2$ and Ti${}_4$O${}_7$, and the interface.}
   \label{fig:potential} 
  \end{center}
 \end{figure}
\end{center}

\section{Band Offset}

After obtaining the three structures - both isolated materials and their interface - it is possible to calculate the natural band offset between the two systems studied in this work. The band offset will be given by \cite{Li2009b}
\begin{align}
 \Delta E_v(\text{TiO}_2|\text{Ti}_4\text{O}_7) &= \Delta E_{v,\phi}(\text{Ti}_4\text{O}_7)-\Delta E_{v,\phi'}(\text{TiO}_2) \nonumber \\ &+\Delta E_{\phi,\phi'}(\text{TiO}_2|\text{Ti}_4\text{O}_7),
 \label{eq:band-offset}
\end{align}
where $\Delta E_{v,\phi}(\text{Ti}_4\text{O}_7)$ and $\Delta E_{v,\phi'}(\text{TiO}_2)$ are the differences between the Valence Band Maximum energy ($E_{VBM}$) and the electrostatic potential averaged over a certain volume of the unit cell of the isolated materials, while $\Delta E_{\phi,\phi'}(\text{TiO}_2|\text{Ti}_4\text{O}_7)$ is the difference between the averaged electrostatic potential of the different materials calculated in bulk-like regions of the relaxed heterostructure supercell.

The determination of the quantities needed by equation \ref{eq:band-offset} can be achieved by, first, obtaining the electronic structure of each isolated (bulk) system. From this calculation, the average electrostatic potential along the crystal axis which is secant to the interface (the $c$ axis, in this case) is determined,
\begin{equation}
 \phi(z) = \frac{1}{ab}\int_0^a \int_0^b u(x,y,z) \, \mathrm{d}x \mathrm{d}y 
 \label{eq:partial}
\end{equation}
as well as the position in energy of the VBM, $E_{VBM}$ for each system. Using the heterostructure after ionic relaxation we also calculated $\phi(z)$ in a similar fashion. From the plot of the electrostatic potential for all three structures (shown in figure \ref{fig:potential}) it is possible to define regions, labeled $\Delta z_1$ and $\Delta z_2$, where both the bulk materials and the heterostructure present similar characteristics. These regions of the heterostructure are considered as bulk-like and should have, in principle, the same potential. The average electrostatic potential is then obtained by the integration of $\phi(z)$ along $\Delta z_1$ and $\Delta z_2$ for Ti${}_4$O${}_7$ and TiO${}_2$ respectively, and is used to calculate the first and second terms of equation \ref{eq:band-offset}. From the heterostructure calculation, the same quantities are used to obtain the final term.

The third step would comprise the determination of the band energy shift due to the hydrostatic pressure. In this particular case our heterostructure is completely strain free, as it is built by equal unit cells, thus this last step presents no contribution. After those steps were carried out, the band offset obtained in our case resulted in $\Delta E_v(\text{TiO}_2|\text{Ti}_4\text{O}_7) = 2.18$ eV. Numerical errors due to the choice of a particular $U$ were estimated to be $\approx$ 0.3 eV, by comparison with other test calculations performed using other values of $U$. Even though the value of the valence band offset presented a variation with respect to the value of $U$ used, being smaller for smaller $U$, the overall electronic structure of the interface has not changed, and the final result, which is the relative position of electronic levels of TiO${}_2$ and Ti${}_4$O${}_7$, remained practically unchanged. For this reason, we have chosen to report only the results for $U$ = 5 eV.

\section{Results}
\label{sec:results}

\begin{center}
 \begin{figure}[Ht!]
  \begin{center}
   \includegraphics[width=\columnwidth]{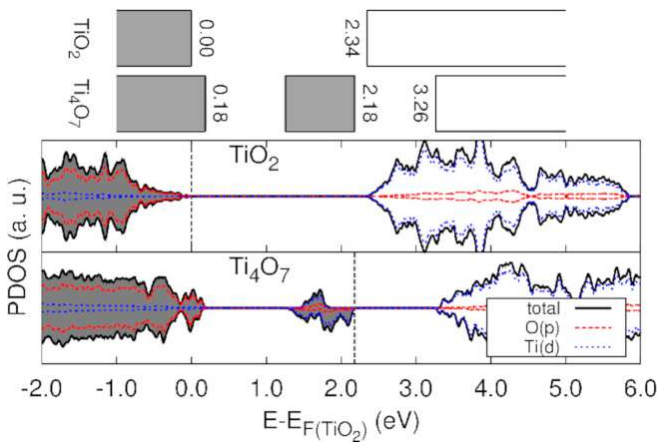}
   \caption{PDOS of TiO${}_2$ rutile and Ti${}_4$O${}_7$. Gray shaded areas denote the occupied levels (below $E_F$, which is highlighted by the dotted vertical lines) and the two spin components are pictured by positive and negative values along the vertical axis. The upper panel shows the band alignment between the two materials.}
   \label{fig:pDOS} 
  \end{center}
 \end{figure}
\end{center}

The projected Density of States (PDOS) was calculated for the two isolated structures and the results were shifted in energy by the band offset. These results are shown in fig. \ref{fig:pDOS}. From that graph, it is possible to notice that the VBM of the interface is located on Ti${}_4$O${}_7$, while the CBM is found in the TiO${}_2$, resulting in a type II heterostructure.

Given that the CBM is 0.16 eV above the VBM (upper panel, Fig. \ref{fig:pDOS}), it is possible that electrons would leave the Ti${}_4$O${}_7$ structure and arrive at the unoccupied levels at the TiO${}_2$ structure due to the presence of an electric field. In this case, the Ti${}_4$O${}_7$ channels present inside of the TiO${}_2$ matrix that forms the memristor could act as donors to the neighboring region of that matrix. This specific characteristic does not depend on the choice of $U$. While it is known that the isolated oxygen vacancy in the TiO${}_2$ rutile acts as a shallow donor,\cite{Janotti2010} as to our knowledge, this same behavior has not been reported for the ordered defects present into the Ti${}_4$O${}_7$ structure.

\begin{center}
 \begin{figure}[t!]
  \begin{subfigure}[FIGTOPCAP,raggedright][ CBM]{
   \includegraphics[width=\columnwidth]{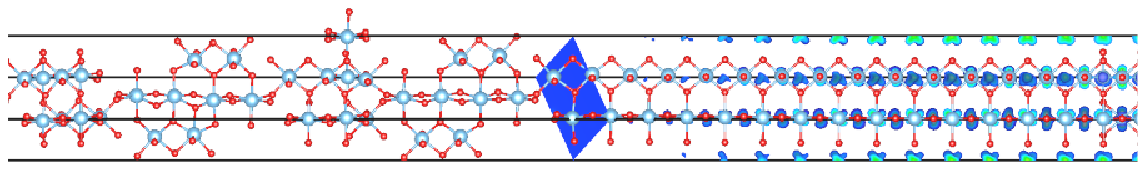}
   \label{fig:cbm}
   }
  \end{subfigure}
  \begin{subfigure}[FIGTOPCAP,raggedright][ VBM]{
   \includegraphics[width=\columnwidth]{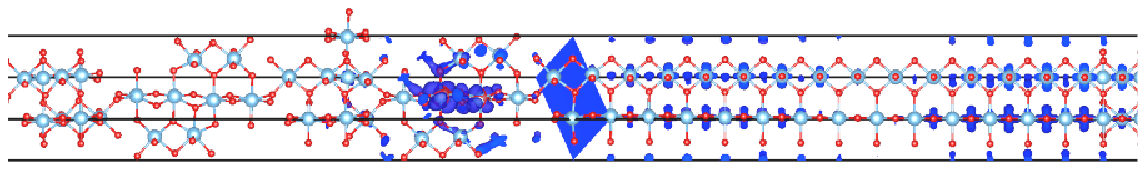}
   \label{fig:vbm}
   }
  \end{subfigure}
  \caption{Projected charge density for the CBM (a) and VBM (b) of the Ti${}_4$O${}_7$-TiO${}_2$ interface after ionic relaxation. The blue plane points out the interface region of the structure, being composed of Ti${}_4$O${}_7$ to the left and TiO${}_2$ to the right of the line.}
  \label{fig:parchg}
 \end{figure}
\end{center}

Further evidence of the real-space separation of the VBM and CBM is given by the projected charge density over the frontier orbitals. The projection was performed for the last occupied and first unoccupied bands, at the $\Gamma$ point. As a result, presented in Fig. \ref{fig:parchg}, it is possible to notice that while the VBM (Fig \ref{fig:vbm}) is located mainly on the Ti atoms of the Ti${}_4$O${}_7$ structure close to the interface (with a contribution of some of the Ti atoms from TiO${}_2$), the CBM (Fig. \ref{fig:cbm}) is located exclusively on the TiO${}_2$ segment. 

Some hybridization may be responsible for the contribution coming from the TiO${}_2$ structure to the VBM, as both levels present Ti(d) character (see Fig \ref{fig:pDOS}). Another explanation for this contribution would be the length of the $c$ axis, which is not long enough to describe the separation of the frontier orbitals. Indeed, calculations performed with a smaller supercell (the supercell $c$ vector length was half of the one used in the calculations presented) resulted in larger charge delocalization. Therefore, we expect that, in the limit of longer unit cells (closer to realistic systems) the charge would be fully separated on the two phases.

\section{Conclusions}

In conclusion, we have reported the band offset given by the average of the electrostatic potential for the Ti${}_4$O${}_7$-TiO${}_2$ heterostructure. The supercell containing the junction was built using transformations that resulted in the same unit cell for both materials present on this junction. The localization of the VBM and CBM was determined to be at the Ti${}_4$O${}_7$ and TiO${}_2$ structures respectively, resulting in a type II heterostructure. As the band gap is 0.16 eV, our results indicate that the Ti${}_4$O${}_7$ channels inside of the memristor TiO${}_2$ matrix would then act as donors for the TiO${}_2$ structure. 

\begin{acknowledgments}
  This work was supported by FAPESP, and CNPq. The support given by Cenapad-SP in the form of computational infrastructure is also acknowledged.
\end{acknowledgments}


\end{document}